# Risk and Machine Protection for Stored Magnetic and Beam Energies

*B. Todd and M. Kwiatkowski*
CERN, Geneva, Switzerland

**Abstract**
Risk is a fundamental consideration when designing electronic systems. For most systems a view of risk can assist in setting design objectives, whereas both a qualitative and quantitative understanding of risk is mandatory when considering protection systems. This paper gives an overview of the risks due to stored magnetic and beam energies in high-energy physics, and shows how a risk-based approach can be used to design new systems mitigating these risks, using a lifecycle inspired by IEC 61508. Designing new systems in high-energy physics can be challenging as new and novel techniques are difficult to quantify and predict. This paper shows how the same lifecycle approach can be used in reverse to analyse existing systems, following their operation and first experiences.

## 1 Introduction

Modern high-energy physics (HEP) machines run with large stored energies. These pose a risk: if energies are released in an uncontrolled manner, there is the potential for damage to accelerator components and an impact on operational availability, as machines have to be stopped and repaired before operations can continue. Risks due to stored energies are studied, and ultimately mitigated by dedicated machine protection systems (MPS). Risk-based approaches should be part of every engineer's toolbox: this is particularly important in the development of systems related to protection. This paper outlines the fundamentals of machine protection, giving an overview of the approaches used in the powering and protection of HEP machines.

### 1.1 The Large Hadron Collider

The Large Hadron Collider (LHC) is the world's most powerful particle accelerator and one of the world's largest and most complicated machines, having been conceived and designed over the course of the last 30 years. This machine represents the cutting edge of accelerator technology. Figure 1 shows a simplified schematic of the LHC, where two counter-rotating beams (Beam-1 and Beam-2) are injected and brought into collision at dedicated experiments located at four of the eight insertion regions (IR) of the machine.

Beams are transferred to the LHC from the Super Proton Synchrotron (SPS), which is the final machine in a complex of accelerators that prepare LHC beams. It takes 12 injections of beam from the SPS to completely fill each of the LHC's two rings. Once filled, the LHC accelerates the beams by means of a radio-frequency (RF) system. The dipole magnetic field is increased in step with the increase in beam energy, ensuring that each circulating beam maintains a consistent beam orbit. Once physics energy is reached, the beams are squeezed and brought into collision. Experiments then gather data, before the machine is emptied, and the process restarted. This process is referred to as the LHC *machine cycle* and is shown in more detail in Fig. 2.

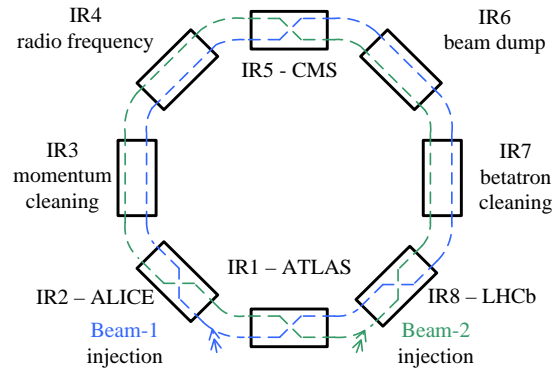

**Fig. 1:** LHC schematic and insertion regions

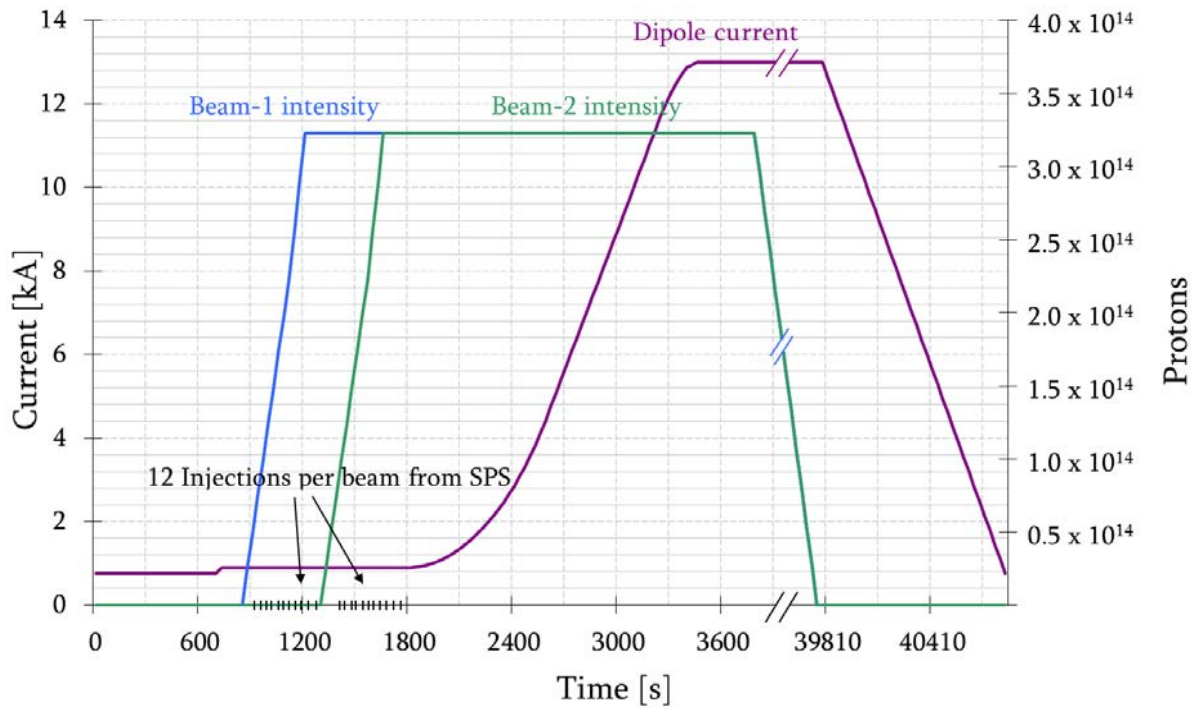

**Fig. 2:** LHC cycle with main dipole current and approximated beam intensities

## 2    Risks due to stored energies

The two principal sources of energy in large physics machines are the powering circuits, in the form of stored magnetic energy, and the beam.

The LHC is designed to provide a centre-of-mass collision energy of 14 TeV ($10^{12}$ eV). This gives around 360 MJ stored beam energy per beam, over 100 times larger than any other machine. To circulate the 7 TeV beams in the 27 km circumference of the machine, magnetic fields of 8.3 T are needed. These fields require superconducting dipole magnets operating at 13 kA, at only around 2°C above absolute zero (about −271°C). At nominal current, around 10 GJ is stored as magnetic energy in the whole magnet powering system. Figure 3 compares the stored magnet and beam energies in the LHC with those of other HEP machines.

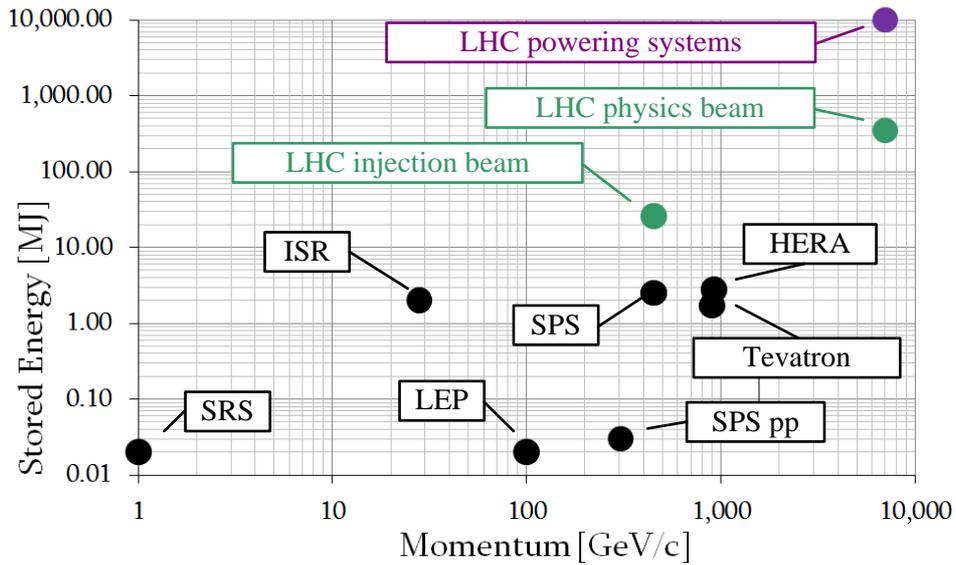

**Fig. 3:** Comparative energy levels in the LHC and other HEP machines [1]

With these parameters, there is a high chance of particle impacts causing a magnet to quench, where superconductor becomes resistive, and action must be taken to prevent damage to the magnets. Losing excessive beam energy in a part of the machine can cause other kinds of damage, such as rupturing the machine vacuum [2]. Table 1 shows the percentages of prompt beam energy losses considered as *quench* and *damage* limits of the LHC.

**Table 1:** Quench and damage levels due to prompt beam energy loss [3]

| Energy [GeV] | Beam energy loss [%] | Consequences |
|---|---|---|
| 450 | 0.000 8 | Quench |
| 450 | 0.5 | Damage |
| 7000 | 0.000 0005 | Quench |
| 7000 | 0.005 | Damage |

In the best case, damage events would result in repairs and downtime; in the worst case there are potential failures that could result in the closure of the laboratory for significant periods of time. One incident regarding an uncontrolled loss of energy from the magnetic circuits has already occurred, and it resulted in CERN's accelerators being closed for over a year to make repairs and to design mitigations. Table 1 represents the key design challenge of machine protection.

At the same time, machines such as the LHC must operate. The performance should be taken as close to operational limits as permitted, whilst at the same time being protected. To accomplish these goals, machine protection systems (MPS) are designed and implemented.

## 3   Machine protection systems (MPS)

At CERN, an MPS has been developed to bring the risks of running the LHC down to an acceptable level. The development process has followed a deep-thinking academic approach as machine protection is not a legal requirement of the LHC project and the MPS is not a safety system, although the two are closely related. In the LHC, the safety system is the LHC Access System, ensuring the collective protection of the personnel against electrical and radiological hazards arising from the LHC accelerator [4].

The MPS is split into two sections, to address the two different classes of risk:

- beam-related machine protection – protecting the LHC against damage due to the accidental release of beam energy, caused by losing beam particles;
- powering machine protection – protecting the LHC against damage due to the accidental release of energy stored in the magnet powering circuits.

Figure 4 shows the relationship between these different systems. In order to correctly protect people and the environment from the risk of operation, the safety system does not require the correct function of the MPS. However, without the MPS, an activation of the safety system would put the LHC machine at risk, as the safety actuators are designed to sacrifice the machine in order to protect the personnel.

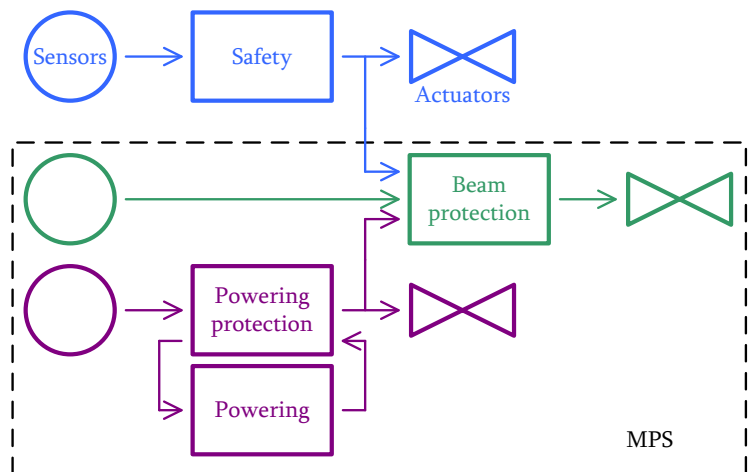

**Fig. 4:** Safety and machine protection system relations

MPS are conceived on the basis of two principles:

- machine safety – the machine must be protected from damage, and should be prevented from unduly wearing out its components;
- machine availability – the machine must provide beams for physics applications.

The *machine safety to availability balance* is a crucial concern in MPS development. A machine that is entirely safe may have low availability, as increasing safety can mean reducing tolerance to non-nominal situations. On the other hand, if the tolerance to non-nominal situations is increased, the safety is generally reduced, whilst the availability is increased.

## 3.1 Non-nominal energy release

Damage to the machine related to energy is triggered by non-nominal energy release: energy releases that cannot be corrected for by using feedback systems or passive protection. Therefore the MPS is based on a combination of two phases.

- *Prevent* non-nominal energy release – intervene in a failure chain that may lead to a non-nominal release. It is known that following the failure of certain components, without intervention, energy release will eventually occur.
- *Protect* equipment from the consequences of non-nominal energy release – intervene if non-nominal energy release occurs, by mitigating the consequences of the release.

## 3.2 Beam energy protections

In the case of non-nominal beam energy loss, the protection phase intervention is performed in three steps, by three types of subsystem, as shown in Fig. 5.

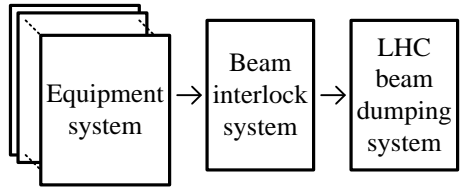

**Fig. 5:** Principal components of the beam-related MPS

- *Equipment systems* are designed to detect non-nominal beam conditions and send beam abort requests to the *Beam Interlock System*;
- The *beam interlock system* transmits the beam abort request to the *LHC beam dumping system*;
- The *LHC beam dumping system* carries out the controlled abort of LHC beam operation, and the dissipation of stored beam energy.

Some failure modes mean that a failure can lead to a dangerous situation in just a few turns of the machine [5]. To protect the LHC from these fast failure modes, the beam-related protection intervention must take place within *several hundred microseconds.*

## 3.3 Powering protection

At design values, the stored energy of all LHC magnet circuits is almost 10 GJ. Powering systems are divided into eight principal subsectors to bring the energy stored in each powering circuit to levels similar to that of other machines [6]. A combination of three key systems performs the intervention in the protection phase intervention as shown in Fig. 6.

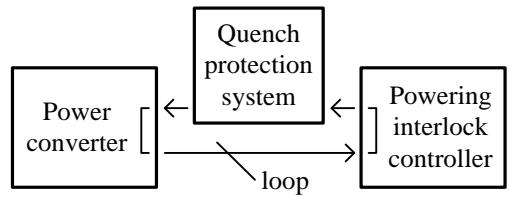

**Fig. 6:** Principal components of the magnet powering-related MPS

- A *quench protection system* detects non-nominal magnet conditions, opens energy extraction switches, and transmits a powering abort request to the power converters.
- The *power converter* stops supplying power to affected circuits.
- In addition, a *powering interlock controller* links related powering circuits, and other subsystems having an influence on LHC magnet powering.

For magnet powering system protection, an intervention must take place within *several milliseconds*.

## 3.4 Protection, safety and control interactions

The safety system is the access system. If intrusion of personnel into the machine environment is detected during operation the access system automatically moves beam stoppers into the LHC beam lines [4]. If circulating LHC beam were to hit these beam stoppers, there would be significant beam losses and the LHC would be damaged. Therefore the access system is connected to both the LHC beam dumping system, and the beam interlock system. If the access system is triggered, the beam-related MPS protects the LHC from the dangers posed by the beam stoppers.

Three of CERN's central systems are also shown in Fig. 7. The *control system* provides general supervision and an overview of all accelerator subsystems. The *timing system* synchronizes all of the accelerator subsystems. Also shown is the safe machine parameters (SMP) system, which is closely related to the MPS. SMP controls the MPS mode; it ensures a consistent configuration of the most critical MPS subsystems. For example, the beam energy losses quench and damage thresholds change as a function of beam energy, as shown in Table 1. SMP broadcasts machine energy throughout the complex, where it is received by, amongst others, the beam loss monitor (BLM) system, and is used to determine whether beam losses are putting the machine at risk.

Interconnections between safety, beam-related machine protection and power protection systems are shown in Fig. 7.

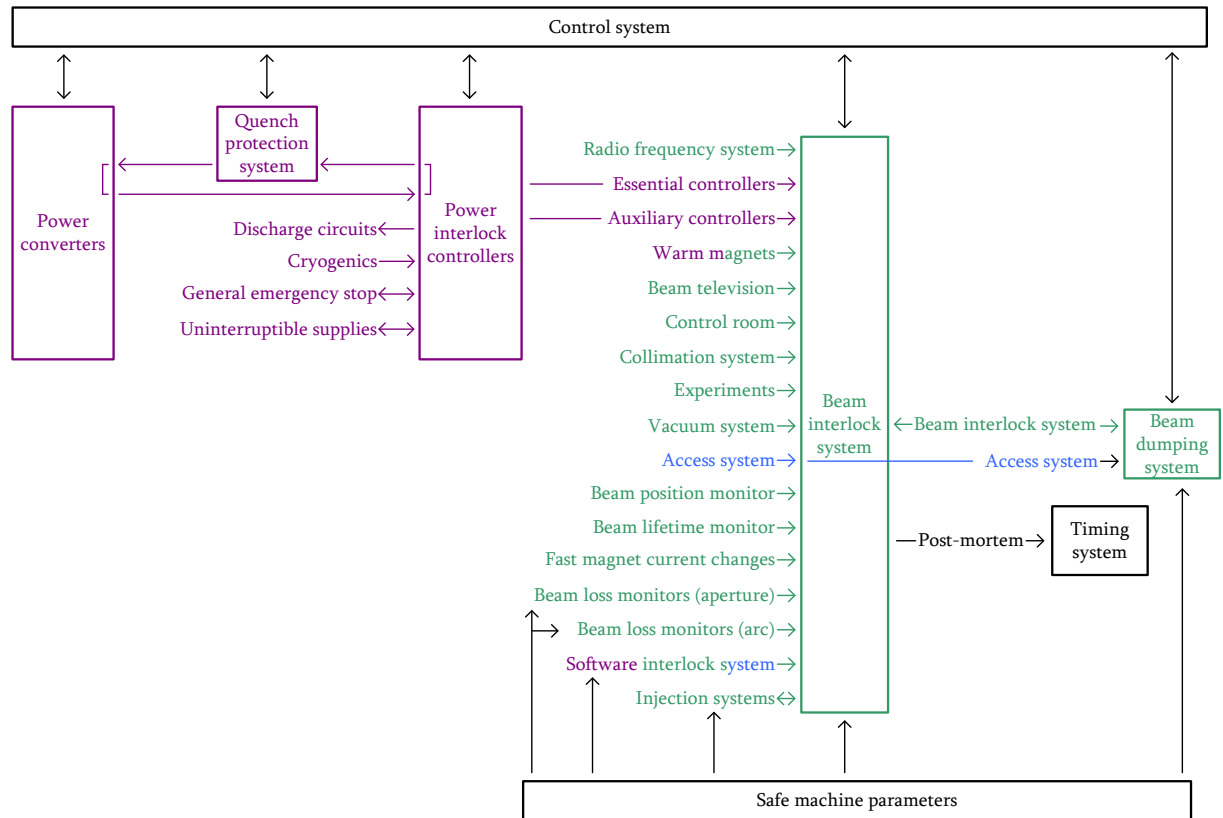

**Fig. 7:** Combined view of safety, protection, and control system interconnections

## 4 Protection system lifecycle: Designing new systems

In order to ensure that risks are mitigated a complete development process needs to be put into place. This ensures that correct decisions are taken at all points during system development. Considering this, the domain of machine protection shares several similarities with the domain of system safety, where MPS is the equivalent to a safety system.

Several industry standards exist, each ensuring that the process of developing safety systems can be proven, and that safety cases can be made. The central standard upon which most modern standards are based is IEC 61508 *Functional Safety of Electrical, Electronic & Programmable Electronic Safety Related Systems* [7], as shown in Fig. 8. IEC 61508 has been used as a basis from which to create the protection system lifecycle, which is outlined in this chapter.

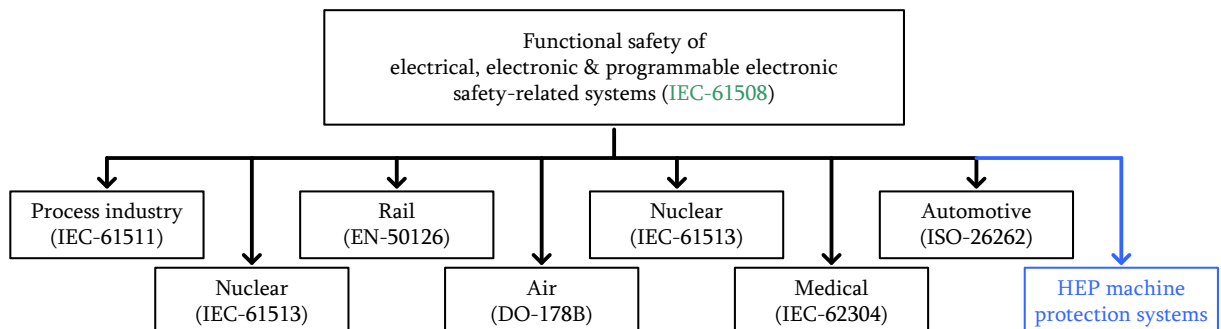

**Fig. 8:** IEC 61508 as a template for other safety standards [8]

With some appropriate conditioning and interpretation, many system-safety techniques, principles, tools, and guidelines can be applied to machine protection. The key steps in the protection system lifecycle are shown in Fig. 9. This is tailored to CERN's academic, non-legal environment, giving a less-rigorous framework than IEC 61508, while establishing best practices and the requirements for MPS development. This process has to be carried out within an organizational structure that is adapted to the role of machine protection. Appendix A outlines the structure currently in place at CERN.

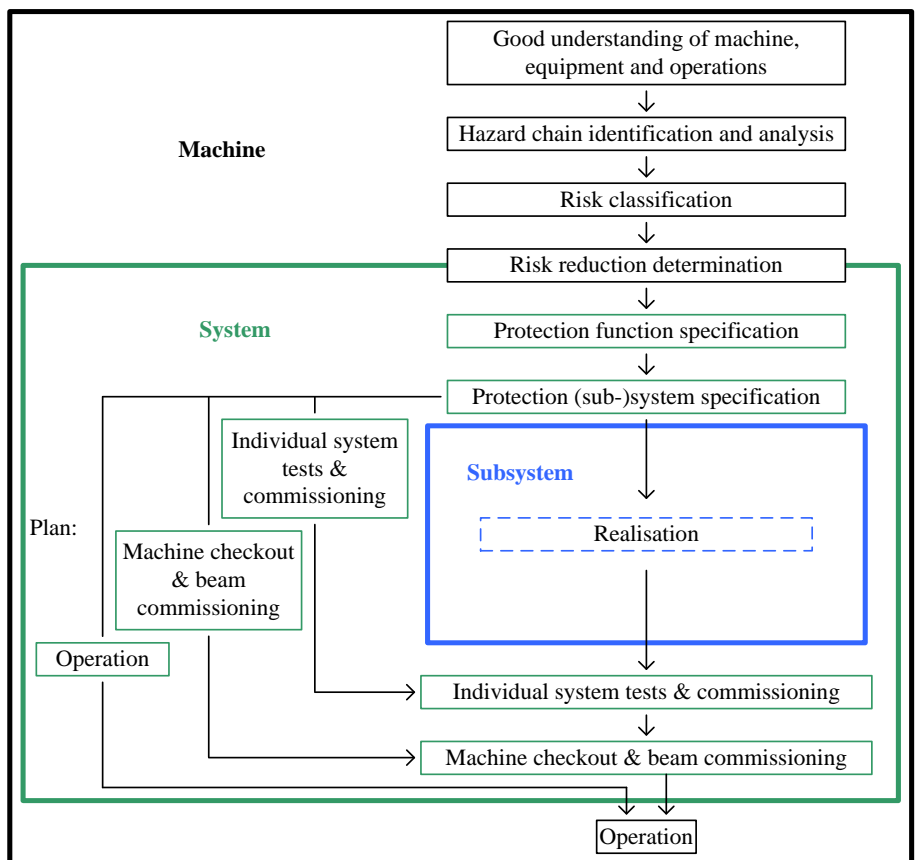

**Fig. 9:** Protection system lifecycle [7]

The first steps of the lifecycle are outlined below, with key information captured while following the process (Fig. 10). This begins with an understanding of risk, and ends with subsystem specifications. The principle is that for each risk identified, it can be proven that adequate mitigation is in place, and that the quality of mitigation is proportional to the risk being mitigated.

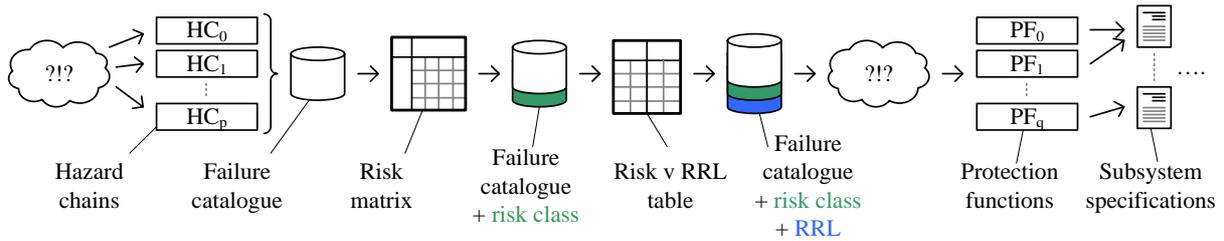

**Fig. 10:** Key data for each step of the protection system lifecycle [7]

### 4.1 Hazard identification and analysis

The first step is to identify *hazard chains*, which link failures to damage through a series of events and/or conditions. The granularity of these depends on the functions being studied, and the types of systems involved in the chain. This step results in the creation of a *failure catalogue*, containing all of the hazard chains that link failures, non-nominal energy deposition, and damage. The template for a failure chain is shown in Fig. 11, with an example explained on the right-hand side.

For each chain, two time constants should be recorded:

– $\Delta T_n$, the time between the failure occurring at the start of the chain and the first non-nominal energy deposition taking place;

– $\Delta T_d$, the time between the first non-nominal energy deposition and damage occurring.

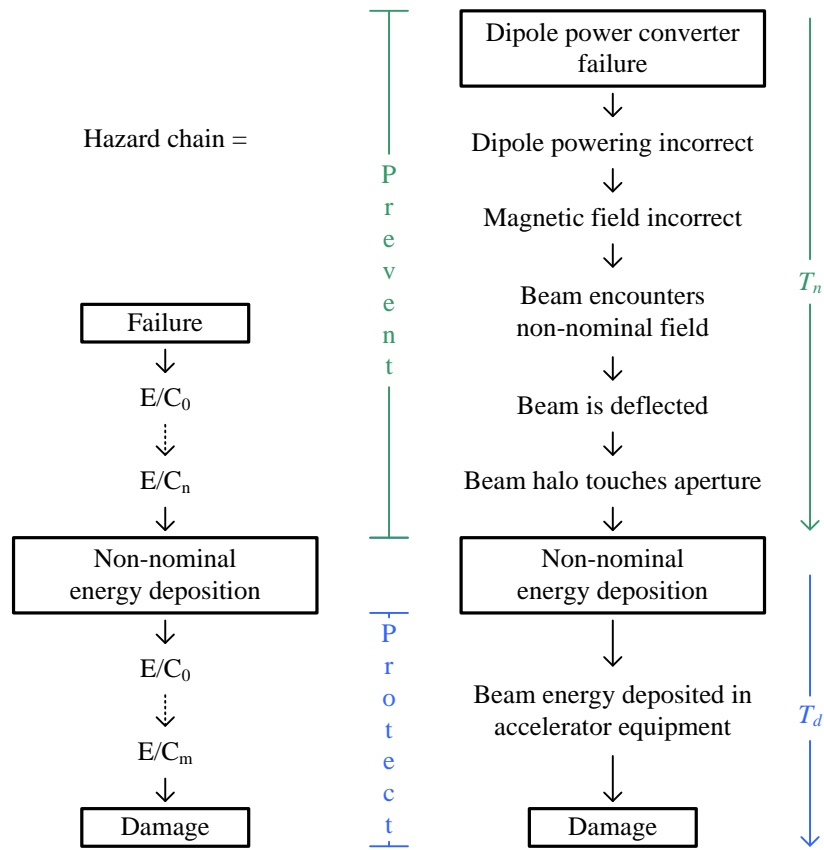

**Fig. 11:** Hazard chains, prevent, protect and time definitions [7]

Chains that have a small total time ($T_n + T_d$) between initial failure and damage can be more difficult to mitigate: less time is available to observe and take action; whereas chains that take a longer overall time to evolve may be more readily observed, and more easily mitigated.

The ratio between $T_n$ and $T_d$ is also important. Failures may take a longer time to lead to non-nominal loss, but then quickly lead to damage, or vice versa. In these cases mitigations must focus on different aspects of the chain, either prevention or protection.

In the example given above, several chains will be created, each for a different magnet failure and magnet families, each having different times between energy deposition and damage.

## 4.2 Risk classification and risk matrix

Risk is defined as a product of *probability* and *impact* [8]. The higher the product, the higher the risk. A *risk matrix* is a visual means of representing this. On one axis is indicated *probability*, and on the other axis is *impact*, with coordinate points representing the individual risks. A typical matrix, with a risk plotted, is shown in Fig. 12.

**Fig. 12:** Risk matrix

The matrix in Fig. 12 is qualitative; this should be converted into a quantitative matrix by appropriate experts. The definition of impact in the case of HEP could be a quantity such as *repair time*, *number of magnets damaged*, or *cost to repair*. In addition to the definition of the axes, the relative risk level for each area should also be qualified, by splitting the matrix into zones of equivalent risk. The typical approach is to choose four or five levels of risk. An example is shown in Fig. 13.

**Fig. 13:** Risk matrix with defined impact and risk levels

A risk level 1 is generally considered 'as low as reasonably possible' (ALARP). When a risk has this level, no mitigation is needed: the risk is accepted. The process of conversion from qualitative to quantitative, and the definition of boundaries, is subjective. There is no standard risk matrix: what is considered as ALARP for one organization or machine may be unacceptable to another.

Each of the hazard chains previously identified should have the associated risk plotted and classified using the risk matrix.

## 4.3 Risk reduction level (RRL)

For each of the risks plotted, a risk reduction level (RRL) is required to bring the risk to the ALARP level: higher risks require a larger RRL (Fig. 14). These RRL are added to the failure catalogue.

| Original | Desired | RRL |
|----------|---------|-----|
| 4 | 1 | 3 |
| 3 | 1 | 2 |
| 2 | 1 | 1 |

**Fig. 14:** Risk reduction level definitions

## 4.4 Protection function specifications

For each hazard chain and associated risk, mitigations must be identified.

The first step is to determine whether there is a means to remove the hazard chain and risk entirely. It may be possible to change the machine architecture to render some chains impossible. If the risk cannot be removed, then *protection functions* should be conceived. These are to intervene in the chains, achieving the required risk reduction levels. A single chain may have several related functions; similarly, a single function may be used to intervene in several chains. If a single chain is broken in multiple places by multiple functions, then the required RRL is shared between the functions.

Functions that affect the HC before non-nominal energy losses occur are *prevention* mechanisms, whereas functions that intervene after non-nominal energy loss, but before damage, are *protection* mechanisms.

## 4.5 Protection integrity level (PIL) specifications

The protection functions are collected, and gathered into technical solutions, which are then assigned to systems. A single system may implement several functions, and a single function may be used to break several hazard chains.

In Fig. 15, hazard chain 1 ($HC_1$) *orbit feedback failure* has three protection functions that intervene to prevent damage. All three functions would have to fail for damage to result. The overall risk is 3, the overall RRL is therefore 2. $PF_1$, $PF_2$, and $PF_3$ must combine to reach this level. However, in hazard chain 2 ($HC_2$) *training quench*, only $PF_3$ interrupts the chain. In this case the risk is 4, giving a RRL of 3. Therefore $PF_3$ *extract energy from powering systems* has to meet the most strenuous requirement of RRL of 3. This is assignment of the so-called *protection integrity level* specification for the protection function.

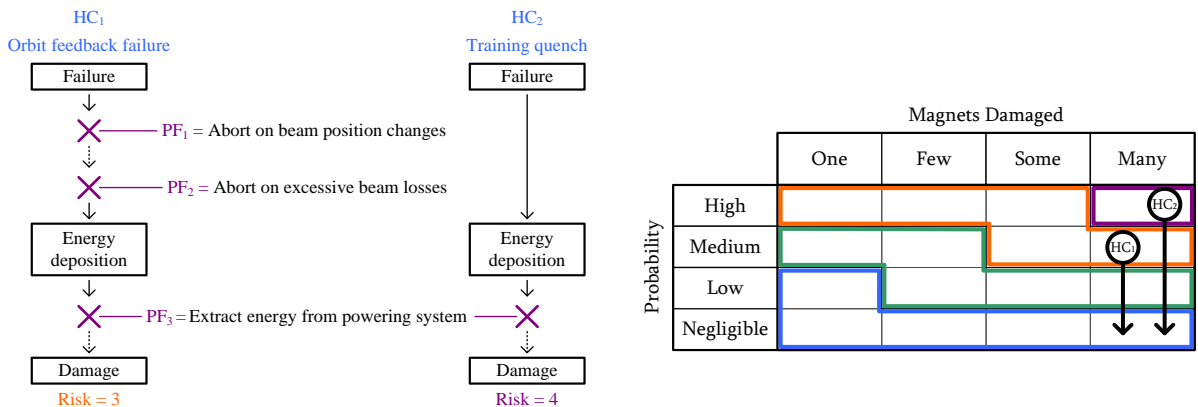

**Fig. 15:** An example of hazard chains and protection functions, with risks and risk matrix

### 4.6 Meeting specifications

The specified protection function has to be realised, and the specified PIL has to be demonstrated; in addition, the subsystem must meet the time constraints $T_n$ and $T_d$ as determined above. The development of high-dependability systems to meet protection requirements forms the central topic of several Ph.D. theses undertaken at CERN, see Appendix B for more information.

### 4.7 Operation and ongoing assurance

The steps following this are installation, testing, commissioning, and finally the handover of the system to operations.

A vital aspect to consider is that the ongoing operation of the machines should always be comparing the observed failure modes and failure rates against what was expected in the hazard chain and risk analysis. For this reason a failure catalogue should be maintained, and each activation of the MPS should be checked to ensure that the MPS is reacting correctly, and that no unforeseen hazard chains occur. If new chains are discovered, the risk analysis is repeated, to build adequate mitigation [9].

## 5 Protection system lifecycle in reverse: Analysing existing systems

The risk-based approach can also be used to *analyse* existing systems and functions. This chapter presents a case study where risk-based analysis is used to consider the existing powering protection in the LHC. The key elements of this discussion are shown in Fig. 16:

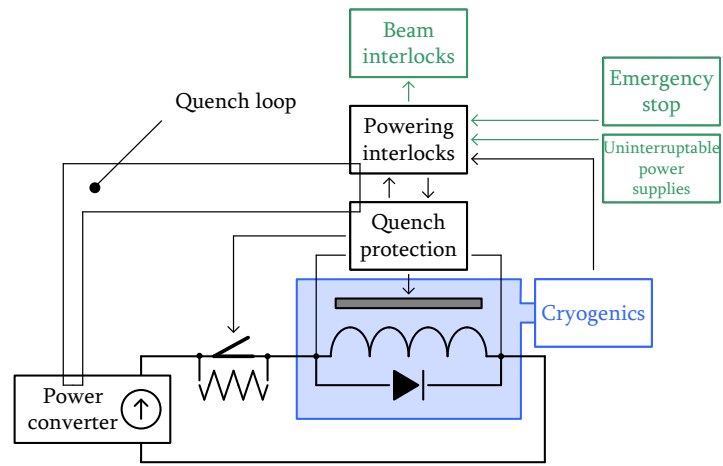

**Fig. 16:** Elements of a superconducting power system with protection

Appendix C gives a detailed description of powering circuit elements. Note that Fig. 16 is a simplification, For example, in the LHC dipole magnet system, 154 dipole magnets are powered in series and it contains two energy extraction systems, one at each end of the circuit.

### 5.1 Non-nominal energy deposition scenario

This case study considers the following electrical scenario, similar to the LHC. In this case seven magnets are connected in three electrical circuits, sharing a cryogenic bath. Consider a non-nominal energy deposition occurring in the central magnet (Fig. 17).

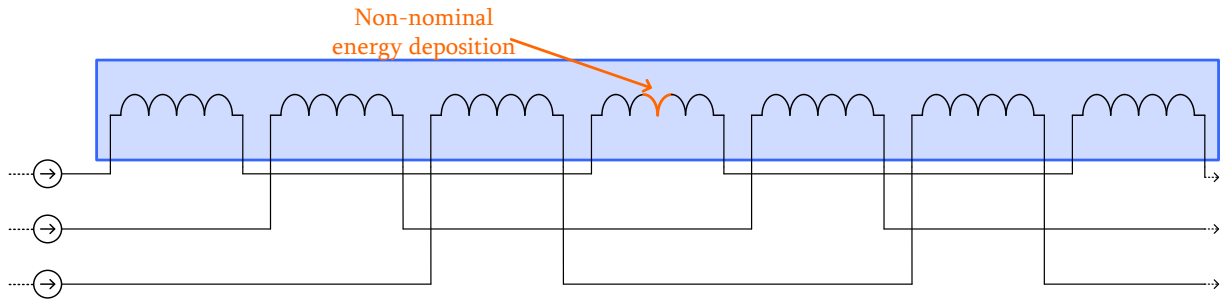

**Fig. 17:** Non-nominal energy deposition scenario

A hazard chain and risk matrix can be constructed using the same approach. In this case, to aid in the discussion, the definition of *probability* and *consequence*, as well as the four *risk classes*, is intentionally vague. The resulting chain and the risk matrix are shown in Fig. 18.

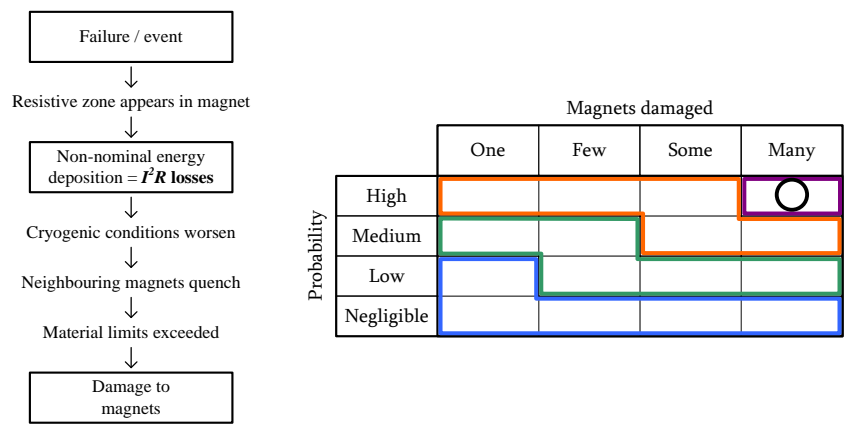

**Fig. 18:** Scenario hazard chain with unmitigated risk

There are four functions in place, breaking the hazard chain (Fig. 19).

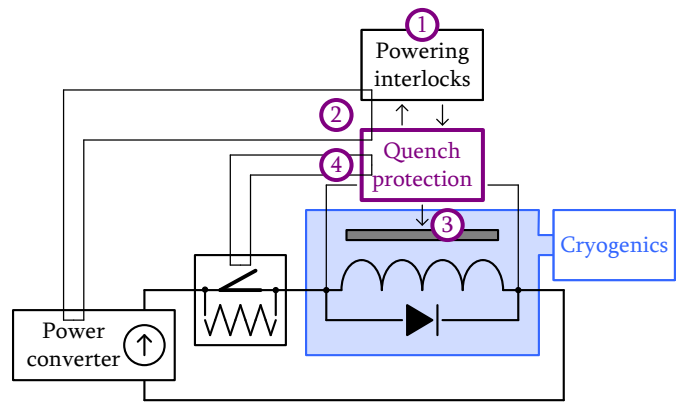

**Fig. 19:** Protection functions in a superconducting circuit scenario

– $PF_1$ – link-related circuits. The quench protection system informs the powering interlock controller of the detection of quench. The powering interlock carries out actions on neighbouring circuits to prevent non-nominal energy deposition in them, which could occur due to quench propagation.

– $PF_2$ – switch off power converter. The power converter is switched off, and no further energy is added to the magnet circuit. The power converter can only slowly extract circuit energy, pushing it back onto the mains supply over the course of several minutes.

- PF$_3$ – propagate the quench. The quench protection system activates a quench heater, which increases the quenched zone in the magnet, increasing the resistive area, and reducing the power density.
- PF$_4$ – extract energy from the magnetic circuit. The quench protection system triggers the opening of the bypass switch where the stored energy of the magnet circuit is removed as heat by including a purpose-built dump resistor in the electrical circuit.

PF$_1$ interrupts the chain in the prevent phase. The other three functions work in the protect phase. On the risk diagram the first function causes the risk to move to the left; as the impact of the failure is reduced, the other protection functions work together to move the risk down, reducing the risk probability (Fig. 20). Hence, the fully mitigated risk, with all protection functions, is in the acceptable region.

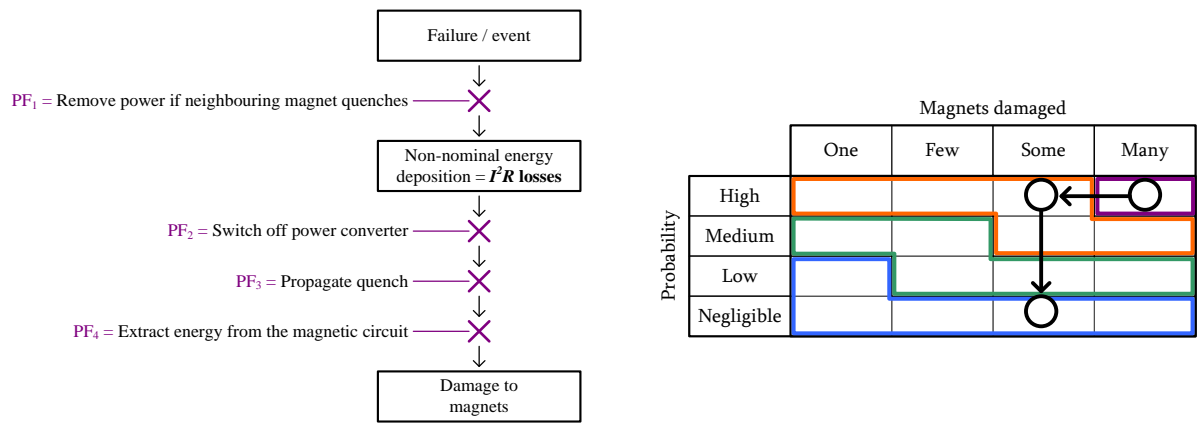

**Fig. 20:** Scenario hazard chain with mitigated risk

## 5.2 Reverse risk analysis

A reverse risk analysis of these protection functions can be carried out by determining the probability and scope of damage following malfunction. The figures in these descriptions come from system experts. The absolute values of risk associated can be debated; this is a worked example of the concept (Fig. 21).

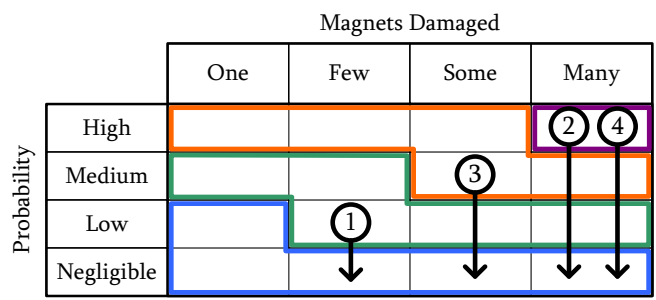

**Fig. 21:** Risk matrix with risk of protection function failure labelled

- PF$_1$ – link related circuits. If neighbouring circuits are not disabled, then the prevent phase of machine protection has been unsuccessful for those circuits; the other functions that are in place in the protect phase are still able to react. This means that there is a low probability of a few magnets being damaged.
- PF$_2$ – switch off power converter. If the power converter does not switch off, it will continue to add power to the quenched circuit until it can no longer maintain the output setpoint. If this occurs, there is a high probability of many magnets being damaged.

- PF$_3$ – propagate the quench. If the quench heater does not fire, the resistive losses in the coil will be localized to a single small area. There is a medium probability that some magnets are damaged.
- PF$_4$ – extract energy from the magnetic circuit. If the energy extraction fails to operate, the field decay time will be very long, vastly increasing the probability of damage occurring; this presents a high probability of many magnets being damaged.

PIL requirements for LHC powering protection systems

Working backwards from the previous risk matrix gives the RRL and consequently the PIL of the protection functions:

- PF$_1$ – link related circuits. Risk class 2 is therefore PIL 1.

  Quench protection system → powering interlock → powering interlock → power converter

- PF$_2$ – switch off power converter. Risk class 4 is therefore PIL 3.

  Quench protection system → power converter

- PF$_3$ – propagate the quench. Risk class 3 is therefore PIL 2.

  Quench protection system → quench heater

- PF$_4$ – extract energy from the magnetic circuit. Risk class 4 is therefore PIL 3.

  Quench protection system → energy extraction

## 5.3 Outcomes of reverse analysis

PIL figures can be compared to system knowledge, identifying elements that require further investigation or, conversely, those systems that have integrity level requirements that are too high. In these cases more methods should be found to understand how to remove the risk, or break the hazard chains more often, so sharing the integrity requirements amongst more systems and functions. This reduces the individual integrity needs of specific systems or connections.

## 6  Conclusions

This paper has described the principal risks in HEP powering systems, the basics of machine protection, the concepts of risk analysis, and a protection system lifecycle.

The paper has shown how the lifecycle can be used in the traditional manner, fitting the concept of IEC 61508, where a protection case is made from first ideas through to system realization. Due to the unique nature of HEP research machines, it is not always possible to be absolutely accurate in the prediction of risks and the assignment of protection functions. With this in mind, the paper concluded with a case study showing that the same lifecycle can be used in a reverse manner to allow analysis and appraisal of existing systems and functions, after machines have been operated and real information is gathered.

Risk is a critical parameter in modern engineering; it should be appreciated and understood by engineers involved in every aspect of HEP. Machine protection is driven by risk analysis. It is a discipline that continues to evolve to meet the changing demands of both the working environment and the types of machines being built. The LHC MPS is novel, being highly tailored to the machine's requirements, having an extremely fast reaction time, and a very high dependability. These are qualities that when combined, cannot be delivered by commercial systems.

## Acknowledgements

This paper has been written in collaboration by the CERN Technology Department power conversion (TE/EPC) and machine protection (TE/MPE) groups. The authors wish to express their gratitude to the numerous individuals who have helped in developing this concept. In addition, the authors wish to acknowledge the contribution of the Warsaw University of Technology, in supervising and reviewing the Ph.D. written on dependable logic design, which first covered the topic of PIL.

## References

[1] Adapted from R. Assmann *et al*., Requirements for the LHC collimation system, Proc. EPAC'02, Paris, 2002, p. 197. http://accelconf.web.cern.ch/AccelConf/e02/PAPERS/TUAGB001.pdf.

[2] R. Schmidt and J. Wenninger. Protection against accidental beam losses at the LHC, Proc. PAC'05, Knoxville, 2005. http://epaper.kek.jp/p05/PAPERS/MOPA005.PDF.

[3] R. Schmidt *et al*., *New J. Phys*. **8**, (2006) 290. http://iopscience.iop.org/1367-2630/8/11/290/.

[4] P. Ninin and L. Scibile, The LHC access system, Proc. EPAC 2004, Lucerne, 2004. http://accelconf.web.cern.ch/AccelConf/e04/PAPERS/WEPLT031.PDF.

[5] A. Gomez Alonso, Barcelona UPC, CERN-THESIS-2009-023, 2009.

[6] M. Zerlauth, Graz University, CERN-THESIS-2005-050, 2004.

[7] IEC, IEC 61508 – Functional safety of electrical/electronic/programmable electronic safety related systems, Part 1, Figure 2, page 17, (IEC, Geneva, 1998).

[8] M. Kwiatkowski, Warsaw University of Technology, CERN-THESIS-2013-216, 2014.

[9] S. Wagner, Zurich ETH, CERN-THESIS-2010-215, 2010.

[10] LHC machine committee website. https://espace.cern.ch/lhc-machine-committee/mission.aspx.

[11] V. Kain, Vienna Tech. U., CERN-THESIS-2005-047, 2005.

[12] A. Vergara, Barcelona Polytechnic U., CERN-THESIS-2004-019, 2003.

[13] B. Todd, Brunel University, CERN-THESIS-2007-019, 2006.

[14] C. Zamantzas, Brunel University, CERN-THESIS-2006-037, 2006.

## Appendix A: Organizational strategy

The development of the MPS is generally carried out by a wide range of experts from multiple disciplines, physicists defining particle simulations and interactions, and system-level experts, working with machine equipment. The MPS traverses organizations, and should be one of the key concerns when considering decisions related to accelerator operation.

It can be observed that organizational structures shown do not reflect a safety management scheme within a company. Structures in HEP are borne out of a confluence of deep-thinking academic and applied industrial approaches. The considerations of the MPS are multi-disciplinary. At CERN an initial machine protection working group (MPWG) was established as a forum to centralize studies, develop scenarios, and to exchange ideas related to machine protection. The MPWG was responsible for coordinating the development of the basic principles of LHC machine protection and organized the production of coherent documentation for all of the major protection systems and subsystems.

The MPWG oversaw the initial commissioning of the LHC powering systems and beam commissioning. Once machine commissioning was completed, the MPWG was renamed to the machine protection panel (MPP) to reflect the changing nature of the LHC from a machine under test to a machine for physics. In addition to this, a subset of the MPP was established, the restricted MPP (rMPP), which consolidates the opinions of the panel members and leads the decision-making processes during operational periods of the machine. Figure A.1 details the background and scale of the members of the MPP and rMPP.

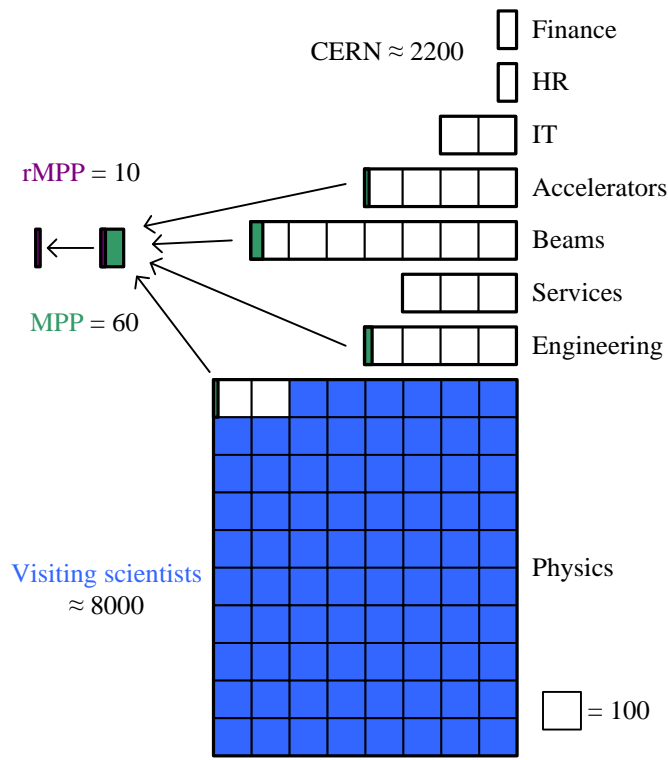

**Fig. A.1:** Protection expertise in CERN

From the machine point of view, the key forum for discussion is the LHC machine committee (LMC). This is an executive committee concerned with technical and performance aspects of the LHC. The LMC is responsible for the definition of operational parameters and the optimization of performance [10]. The rMPP reports on behalf of the MPP to the LMC. From this, and other reporting groups, senior management has sufficient information to take informed decisions. In this way there is a clear chain of command concerning machine protection issues (Fig. A.2): from the frontline engineering teams, through to senior management.

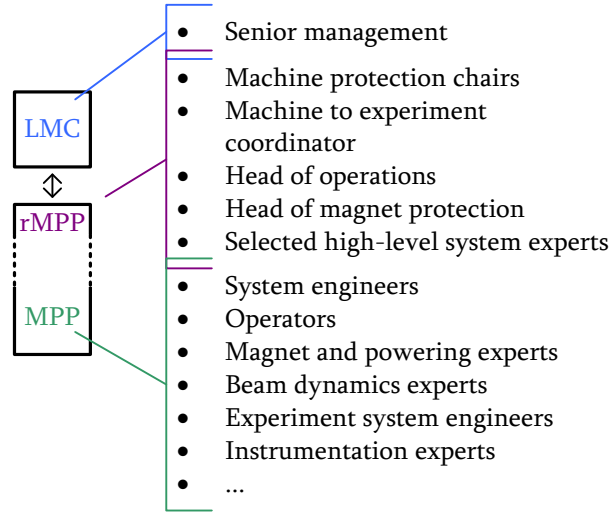

Fig. A.2: Protection chain of command

## Appendix B: MPS doctoral research at CERN

MPS elements highlighted in Figure B.1 are those that have been subjects of doctoral level research.

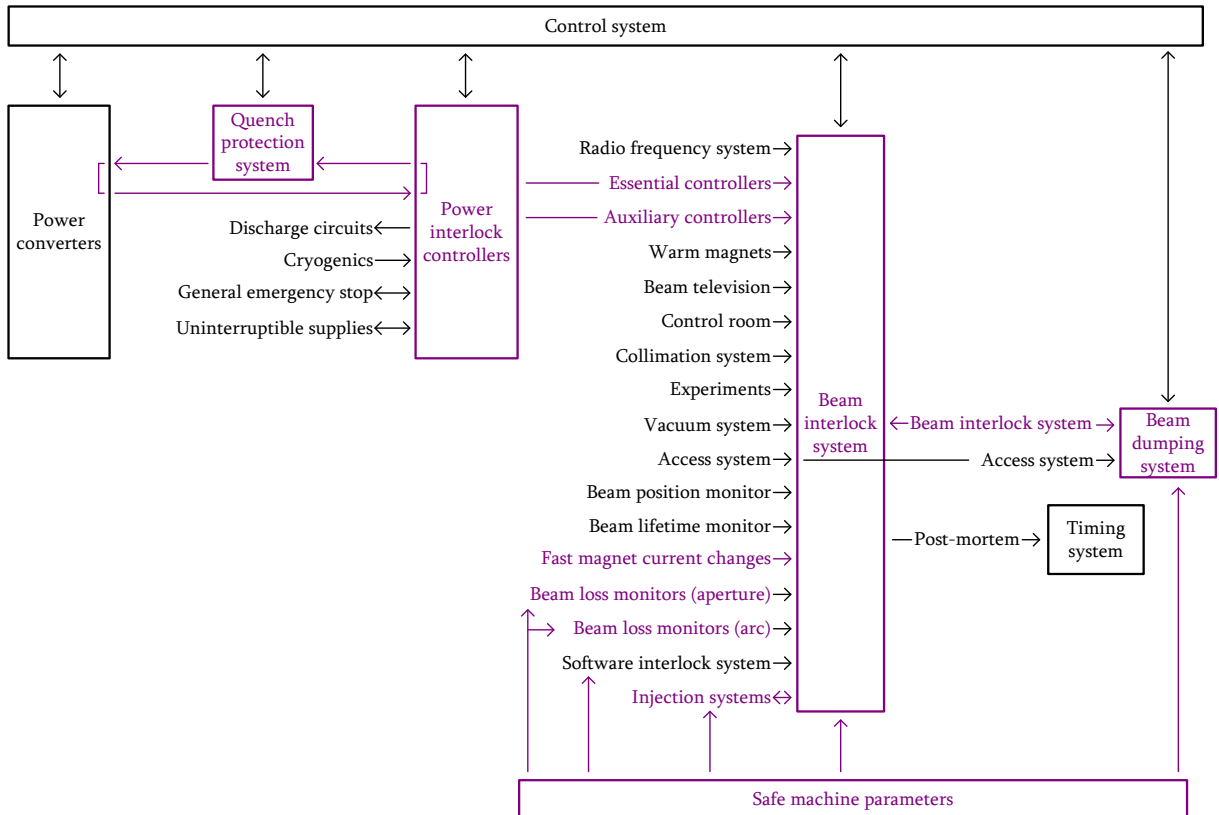

Fig. B.1: Elements of the MPS subject to Ph.D. research at CERN

- *Injection systems* – Machine protection & beam quality during the LHC injection process, V. Kain [11].

- *Quench protection system* – Reliability of the quench protection system for the LHC superconducting elements, A. Vergara [12].
- *Fast magnet current change monitor* – Redundancy of the LHC machine protection systems in case of magnet failures A. Gomez Alonso [5].
- *Beam interlock system* – A beam interlock system for CERN high energy accelerators, B. Todd [13].
- *Safe machine parameters* – Methods for the application of programmable logic devices in electronic protection systems for high energy particle accelerators, M. Kwiatkowski [7].
- *Power interlock controllers* – Powering and machine protection of the superconducting LHC accelerator, M. Zerlauth [6].
- *Beam loss monitor system* – Reliability of the beam loss monitors system for the Large Hadron Collider at CERN, G. Guaglio, and The real-time data analysis and decision system for particle flux detection in the LHC accelerator at CERN, C. Zamantzas [14].
- *Safety vs. availability* – LHC machine protection system: Method for balancing machine safety and beam availability, S. Wagner [9].

## Appendix C: Elements of superconducting powering systems and protection

The key elements of the powering and protection of superconducting circuits are shown in Fig. C.1.

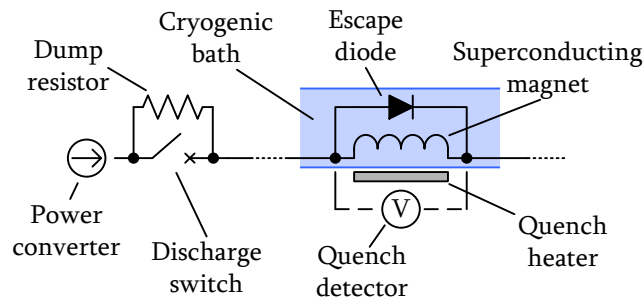

**Fig. C.1:** Components of a superconducting magnet and protection system

- Power converter, superconducting magnet, and cryogenic bath – these form the equipment under control. The converter regulates the magnet current required to achieve the machine beam performance requirements. The cryogenic system, of which only the bath is shown above, is responsible for maintaining the low temperatures required by the magnet in order to achieve superconductivity.
- Quench detector – this element determines the resistance of the magnet coil, by measuring the voltage drop from one end of the magnet to the other. An ideal superconductor should have no voltage drop, as it has no internal resistance. A voltage appearing over the magnet coil is an indicator that a quench is occurring, and is used to initiate the powering protection functions.
- Quench heater – once a quench is detected, heaters are used to force the quench to propagate over a large area. This reduces the power density of the quench, and decreases the chances of damage to the magnet coil.
- Escape diode – this further protects the magnet coil by providing an alternative current path around the quenching magnet coil. As the magnet quenches, the voltage drop increases, and at one point this becomes high enough to switch on the escape diode. From this point the current bypasses the quenching magnet coil.

- Discharge switch and dump resistor – stored magnetic energy in the superconducting circuit has to be discharged after a quench has occurred. This is carried out by opening a discharge switch and adding a dump resistor in series with the magnet coil; the current in the circuit diminishes as power is lost as heat in the dump resistor.